\documentclass[runningheads,a4paper]{llncs}
\pdfoutput=1
\usepackage{enumitem}
\usepackage{graphicx}
\usepackage{amsmath}
\usepackage{amsfonts}
\usepackage{amssymb}
\usepackage{verbatim}
\usepackage{amssymb}
\usepackage{mathtools}
\usepackage{graphicx}
\usepackage{comment}
\usepackage{color}

\input{Qcircuit.tex}
\def\beq{\begin{equation}}
\def\eeq{\end{equation}}
\def\beqa{\begin{eqnarray}}
\def\eeqa{\end{eqnarray}}

\usepackage{url}
\urldef{\mailsa}\path|{alfred.hofmann, ursula.barth, ingrid.haas, frank.holzwarth,|
\urldef{\mailsb}\path|anna.kramer, leonie.kunz, christine.reiss, nicole.sator,|
\urldef{\mailsc}\path|erika.siebert-cole, peter.strasser, lncs}@springer.com|
\newcommand{\keywords}[1]{\par\addvspace\baselineskip
\noindent\keywordname\enspace\ignorespaces#1}

\usepackage{algorithm}
\usepackage{algorithmic}
\floatname{algorithm}{Protocol}

\begin{document}

\mainmatter

\title{An Evolutionary Approach to Optimizing Communication Cost in Distributed Quantum Computation}

\titlerunning{An Evolutionary Approach to Optimizing Communication Cost in Distributed Quantum Computation}

\author{Mahboobeh Houshmand$^{1}$
\and Zahra Mohammadi$^{2}$\and Mariam Zomorodi-Moghadam$^{3}$\and Monireh Houshmand$^{2}$
}

\authorrunning{Mahboobeh Houshmand, Zahra Mohammadi, Mariam Zomorodi-Moghadam, and Monireh Houshmand}

\institute{$^{1}$Department of Computer Engineering, Mashhad Branch, Islamic Azad University, Mashhad, Iran\\
$^2$Department of Electrical Engineering, Imam Reza International University, Mashhad, Iran\\
$^3$Department of Computer Engineering, Ferdowsi University of Mashhad, Mashhad, Iran
}

\toctitle{}
\tocauthor{}
\maketitle

\begin{abstract}
Distributed quantum computing has been well-known for many years as a system composed of a number of small-capacity quantum circuits.
Limitations in the capacity of monolithic quantum computing systems can be overcome by using distributed quantum systems which communicate with each
other through known communication links. In our previous study, an algorithm with an exponential complexity was proposed to optimize the number of qubit
teleportations required for the communications between two partitions of a distributed quantum circuit. In this work, a genetic algorithm is used to solve
the optimization problem in a more efficient way. The results are compared with the previous study and we show that our approach works almost the same
with a remarkable speed-up. Moreover, the comparison of the proposed approach based on GA with a random search over the search space verifies the effectiveness of GA.
\end{abstract}

\section{Introduction}

Quantum computing is one of the emerging technologies in computation with a great potential to outperform classical computers~\cite{shor-1997-26,grover1996fast,grover1997quantum}. The theory of quantum computing is getting more and more mature since it was initiated by Feynman and Deutsch in the 1980s. Quantum computation has revolutionized the computer science, showing that the processing of quantum states can lead to a remarkable speed up in the solution of a class of problems, as compared to the traditional algorithms that process classical bits.

One of the implementation limitations of quantum computing is due to the interactions of qubits with the environment. When the number of qubits increases, the quantum information becomes more fragile and more susceptible to errors~\cite{stolze2008quantum}. On the other hand, error-correction codes require the involvement of many qubits just for one logical qubit and so a number of logical qubits may not fit on a single quantum chip~\cite{van2010distributed}. To overcome this problem, a distributed quantum system is a reasonable solution in which fewer qubits are used in each node or subsystem. Therefore, to have a large quantum computer, one appropriate solution is to build a network of finite quantum computers which are interconnected through a quantum or classical channel and they can implement the behavior of the whole quantum system. This structure is known as ``distributed quantum computing"
~\cite{van2010distributed}. It should be noted that distributed quantum computing in the sense implied in this study is different with anything that is ``distributed and quantum", e.g., quantum key distribution protocols~\cite{bennett1984quantum,ekert1991quantum} in which a secret key is classically distributed between two distant parties.

The circuit model for quantum computation can be expanded to the case of distributed quantum computation where each subsystem sends its data on demand to the other sites through a communication channel. A reliable mechanism for such communication is by using quantum teleportation between nodes of a distributed quantum system.

Quantum teleportation is one of the most promising ways for quantum state movement and its validity has been demonstrated experimentally in some studies including~\cite{furusawa1998unconditional,bouwmeester1997experimental,metcalf2014quantum}. The basic idea in the quantum teleportation is to transfer qubit states without moving them physically. The original state is destroyed such that quantum teleportation does not contract with the no-cloning theorem. The no-cloning theorem states that it is impossible to produce an identical copy of a qubit which is initially in an arbitrary state. Quantum teleportation is composed of some phases including creating an EPR pair, doing some local operations, measurement, and classical communication~\cite{van2007communication}.

In the distributed realization, communication can be done by teleporting qubits between communication links. Although quantum circuit for teleportation is a much simpler circuit compared to any nontrivial quantum computational task~\cite{bashar2009review}, having many such teleportation circuits, maybe a thousand, in interconnection links results in high communication costs for distributed quantum circuits. A non-functional property~\cite{troya2009specification} of a distributed quantum system, which is a constraint on the manner the system performs its functionality, is its efficiency. Efficiency is a quality that reflects a system's ability to meet its performance requirements while minimizing its usage of resources. In this study, minimizing the number of teleportations between nodes of a distributed quantum computer is considered as a measure of its efficiency.

In our previous study~\cite{zomorodi2018optimizing}, with two given quantum subsystems, an algorithm with an exponential complexity was proposed to optimize the number of qubit teleportations required for the communication between these two subsystems. In this work, the genetic algorithm is used to solve the optimization problem in a more efficient way.

In Section~\ref{sec:background}, the required background is given. The related work of the distributed quantum computing is presented in Section~\ref{sec:related work}. Our proposed approach and the results of the work are explained in Sections~\ref{sec:proposed} and~\ref{sec:results}, respectively. We conclude the paper in Section~\ref{conclusion}.

\section{Background}
\label{sec:background}
In this section, the required background is introduced using quantum teleporation as an example.

The basic idea in quantum teleportation is to transfer an unknown quantum
state of a quantum bit (qubit) using two classical bits in a way that the receiver reproduces exactly the same state as the original qubit state. The original state is destroyed such that quantum teleportation does not contract with the no-cloning theorem. The no-cloning theorem~\cite{Nielsen10} is one of the earliest results of quantum computation and quantum information which states that an unknown quantum
system cannot be cloned by unitary transformations.

Before proceeding with the description of the quantum teleportation, Diract notation, qubits, quantum gates, and quantum circuits are introduced.
 In quantum mechanics, Diract or bra–ket notation is a standard notation for describing quantum states, where kets like $\left\vert v\right\rangle$ are column vectors. The bra vector $\left\langle v \right|$ is a row vector and the conjugate transpose of $\left\vert v\right\rangle$. The basis states, $\left\vert 0\right\rangle, \left\vert 1\right\rangle$, are identified by $\left[ \begin{array}{l}1 \\0 \\\end{array} \right]$, $\left[ \begin{array}{l} 0 \\ 1 \\ \end{array} \right]$, respectively.

 \emph{Qubits} are quantum analogues of classical bits. A qubit is a two-level quantum system whose state is represented by a unit vector in a two-dimensional complex space, for which an orthonormal basis, denoted by $\{\left\vert 0\right\rangle$, $\left\vert 1\right\rangle\}$, has been fixed.
Unlike classical bits, qubits can be in a superposition of $\left\vert 0\right\rangle$ and $\left\vert 1\right\rangle$ which is represented as $\alpha\left\vert 0\right\rangle+\beta\left\vert 1\right\rangle$, where $\alpha$ and $\beta$ are complex numbers such that
$|\alpha|^2 + |\beta|^2 = 1$. If such a superposition is measured with respect to the $\{$$\left\vert 0\right\rangle$, $\left\vert 1\right\rangle$$\}$ basis, then the classical outcome of 0 is observed with the probability of $|\alpha|^2$ and the classical result of 1 is observed with the probability of $|\beta|^2$. If 0 is obtained, the state of the system after the measurement will collapse to $\left\vert 0\right\rangle$ and if 1 is obtained, it will be $\left\vert 1\right\rangle$.

The circuit model for quantum computation is based on unitary evolution of qubits by networks of gates~\cite{Nielsen10}. Every quantum gate is a linear transformation represented by a unitary matrix. A matrix $U$ is \emph{unitary} if $UU^{\dagger} = I$, where $U^{\dagger}$ is the conjugate transpose of the matrix $U$.
Some useful single-qubit gates are the elements of the Pauli set:
\[
\sigma_0=
I=%
\begin{bmatrix}
1 & 0\\
0 & 1
\end{bmatrix},
\sigma_1=
X=%
\begin{bmatrix}
0 & 1\\
1 & 0
\end{bmatrix}, \]
\[
\sigma_2=
Y=%
\begin{bmatrix}
0 & -i\\
i & 0
\end{bmatrix},
\sigma_3=
Z=%
\begin{bmatrix}
1 & 0\\
0 & -1
\end{bmatrix}.
\]

Hadamard, $H$, and $R_k$ are other known single-qubit gates where:
\[
H=%
\frac{1}{\sqrt{2}}\begin{bmatrix}
1 & 1\\
1 & -1
\end{bmatrix},
R_k=
\begin{bmatrix}
1 & 0\\
0 & e^{\frac{2\pi i }{2^k}}\,
\end{bmatrix}
.
\]

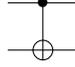
\begin{figure}
\centering
\[
\Qcircuit @C=1.0em @R=1.4em{
& \ctrl{1}&\qw\\
& \targ&\qw
}
\]
\caption{The circuit representation of CNOT gate.}
\label{fig:cnot}
\end{figure}

If $U$ is a gate that operates on a single qubit, then controlled-$U$ is a gate that operates on two qubits, i.e., control and target qubits, and \emph{U} is applied to the target qubit if the control qubit is $\left\vert 1\right\rangle$ and leaves it unchanged otherwise. For example, controlled-NOT (CNOT) gate performs the $X$ operator on the target qubit if the control qubit is $\left\vert 1\right\rangle$. Otherwise, the target qubit remains unchanged.
Figure~\ref{fig:cnot} shows the circuit representation of the CNOT gate.

A quantum circuit consists of quantum gates interconnected by quantum wires carrying qubits with time flowing from left or right. The unitary matrix of the quantum circuit is evaluated by either matrix product or tensor product~\cite{Nielsen10} of the unitary matrices of those quantum gates. The net effect of the gates which are applied to the same subset of qubits in series is computed by the matrix multiplication. The adjacent gates which act on independent subsets of qubits can be applied in parallel and their overall net effect is computed by their tensor product. To realize arbitrary quantum gates, they are decomposed to a set of physically implementable gates by quantum technologies (typically CNOT and single-qubit gates, called``basic gate" library~\cite{Barenco95}), which is called quantum logic synthesis~\cite{mottonen2004quantum,Shende06,zomorodi2014synthesis,zomorodi2016rotation,houshmand2014,houshmand2015}.

Basically, in quantum teleportation, two parties, referred to as Alice and Bob, share an entangled pair of qubits, e.g., $\left| {\beta_{00} } \right\rangle $ (defined in~(\ref{bell})).
\emph{Entanglement} is a quantum mechanical phenomenon that plays a key role in many of the most interesting applications of quantum computation and quantum information.
A multi-qubit quantum state $\left\vert v \right\rangle$ is said to be entangled if it cannot be written as
the tensor product $\left\vert v \right\rangle=\left\vert \phi_1 \right\rangle \otimes \left\vert \phi_2\right \rangle $ of two pure states. For example, the EPR pair~\cite{Nielsen10} shown below is an entangled quantum state:
\begin{equation}
\left| {\beta_{00} } \right\rangle=(\left\vert 00 \right\rangle+\left\vert 11 \right\rangle)/\surd{2}
\label{bell}
\end{equation}
Alice attempts to send an unknown qubit, $ \left| \psi  \right\rangle $, to Bob.
The overall state of the system $(\left| \phi  \right\rangle)$ is as follows:
\[
\begin{array}{l}
\left| \phi  \right\rangle  = \left| \psi   \right\rangle  \otimes \left| {\beta _{00} } \right\rangle  = (\alpha \left| 0 \right\rangle  + \beta \left| 1 \right\rangle ) \otimes (\frac{{\left| {00} \right\rangle  + \left| {11} \right\rangle }}{{\sqrt 2 }}) =  \\
  \frac{1}{\sqrt 2}( {\alpha (\left| {000} \right\rangle  + \left| {011} \right\rangle ) + \beta (\left| {000} \right\rangle  + \left| {111} \right\rangle )}) \\
 \end{array}
\]
The first two qubits belong to Alice and Bob has the third qubit. Alice applies a CNOT gate to the first two qubits and then she applies a Hadamard gate to the first qubit which results in:
\[
\begin{array}{l}
 \frac{1}{2}[a(\left| {000} \right\rangle  + \left| {011} \right\rangle  + \left| {100} \right\rangle  + \left| {111} \right\rangle ) + b(\left| {010} \right\rangle  + \left| {001} \right\rangle  - \left| {110} \right\rangle  - \left| {101} \right\rangle )] \\
  = \frac{1}{2}[\left| {00} \right\rangle (a\left| 0 \right\rangle  + b\left| 1 \right\rangle ) + \left| {01} \right\rangle (a\left| 1 \right\rangle  + b\left| 0 \right\rangle ) + \left| {10} \right\rangle (a\left| 0 \right\rangle  - b\left| 1 \right\rangle ) + \left| {11} \right\rangle (a\left| 1 \right\rangle  - b\left| 0 \right\rangle )] \\
 \end{array}
\]

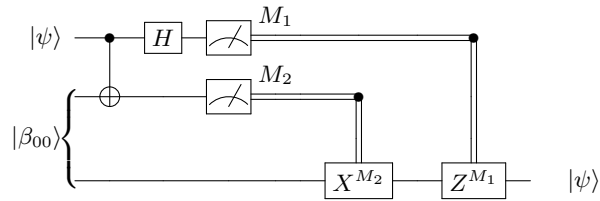
\begin{figure}[ht]
\centering
\[
\Qcircuit @C=1em @R=1em{
\lstick{\ket{\psi}} & \ctrl{1} & \gate{H} &\meter &\ustick{M_1} \cw  & \cw & \cw & \cw & \control \cw & & \\
 & \targ & \qw & \meter & \ustick{M_2} \cw & \cw &  \control \cw & &\cwx & & \\
\lstick{\ket{\beta_{00}}} & & & & & & \cwx & &\cwx & & \\
 & \qw &\qw & \qw & \qw & \qw & \gate{X^{M_2}} \cwx & \qw & \gate{Z^{M_1}} \cwx & \qw & \rstick{\ket{\psi}} \gategroup{2}{1}{4}{1}{.5em}{\{}
}
\]
\caption{Quantum circuit for teleporting a qubit~\cite{Nielsen10}. The two top lines belong to Alice, while
Bob has the third qubit. The meters represent quantum measurement, and the double lines coming out of them carry classical
bits.}
\label{fig:tele}
\end{figure}

Finally, Alice measures both qubits that belong to her. She will obtain one of the four outcome states of $\left| {00} \right\rangle$, $\left| {01} \right\rangle$, $\left| {10} \right\rangle$ or $\left| {11} \right\rangle$ with the equal probability of $\frac{1}{4}$. Depending on the result of Alice's measurement, Bob's qubit collapses to $a\left| 0 \right\rangle  + b\left| 1 \right\rangle$, $a\left| 1 \right\rangle  + b\left| 0 \right\rangle$,
$a\left| 0 \right\rangle  - b\left| 1 \right\rangle$ or $a\left| 1 \right\rangle - b\left| 0 \right\rangle$, respectively. Alice then sends the results of her measurement to Bob using two classical bits. Alice's initial qubit, $ \left| \psi  \right\rangle $ is totally destroyed upon her measurement which makes the quantum teleportation consistent with the no-cloning theorem. Finally, after receiving the two classical bits, Bob can know the state of the qubit at his hand by applying $I$, $X$, $Z$ and $Y$, if the classical bits are 00, 01, 10 and 11 respectively to reconstruct the initial state of Alice.
% of Here a classical communications channel is required. Alice announces the results of her measurements to Bob. Based on them, Bob transfers the state of his qubit into Alice's unknown state by applying the appropriate gate on it.
Figure~\ref{fig:tele} shows the circuit for the entire teleportation protocol as stated above.

\begin{figure}
\centering
    \includegraphics[width=\textwidth]{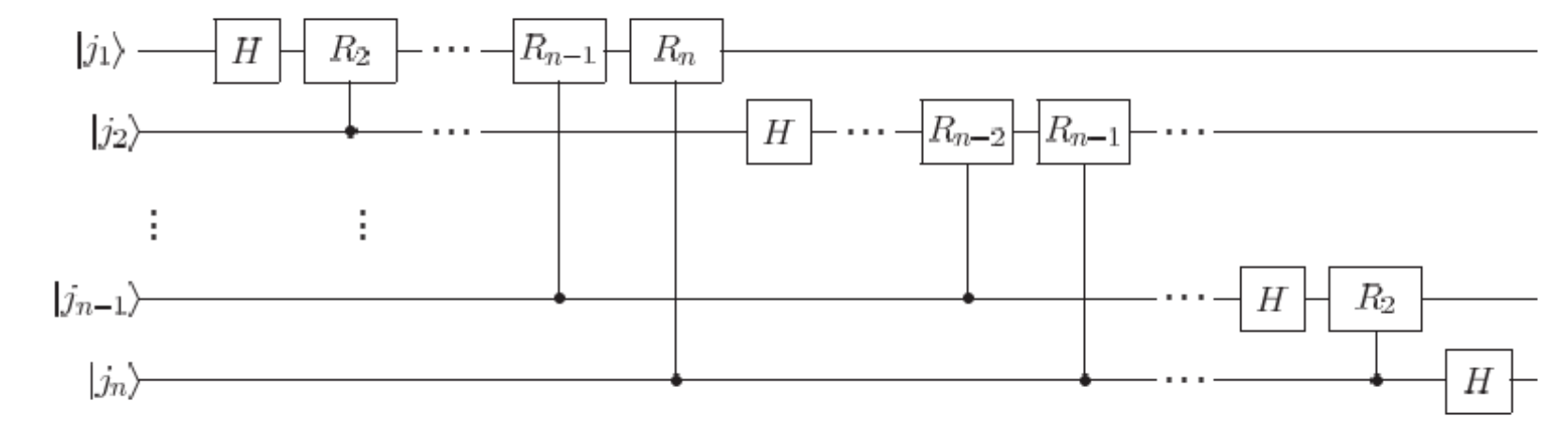}
  \caption{QFT Circuit}
  \label{fig:fig2}
\end{figure}

Another important quantum circuit which is also one of the components of Shor's algorithm \cite{shor-1997-26} is Quantum Fourier Transform (QFT). As Shor's algorithm has potential application in breaking the RSA encryption protocol, implementation of this algorithm has been extensively investigated in the literature. Figure~\ref{fig:fig2} shows QFT for an $n$-qubit state.

\section{Related Work}
\label{sec:related work}
Evolutionary algorithms have been successfully used in different aspects of quantum computing. For example, genetic programming can be used to evolve quantum programs and algorithms~\cite{spector1999quantum}. Having a simulator for a quantum computer, genetic programming can be used to calculate the fitness of a program in the population on the simulated quantum computer. In~\cite{stepney2008searching}, a review on
how evolutionary algorithms have been used to evolve quantum circuits is given. In that study, issues in representing quantum artefacts
in a form suitable for evolutionary search are discussed and it is stated that the basic approach  uses bit strings  to encode the search space and quantum gates. Moreover, different quantum artefacts
that have been discovered through evolutionary search, an example of which QFT, are presented.
In~\cite{leier2004comparison}, genetic programming has been used to find the best quantum circuit for a given quantum algorithm. In that study, the authors have tested different selection strategies for the evolutionary quantum circuit design and showed that the tournament selection and the self-adaptation have been effective on the test problems. The approach in~\cite{massey2004evolving} uses a new technique to evolve a quantum circuit, different from~\cite{leier2004comparison} and the presented programs and circuits were capable of solving specific problem instances and one general problem.
Also in another work~\cite{spector2008machine}, a new quantum circuit has been evolved for the two-query AND/OR problem.
An efficient approach based on genetic algorithms is proposed to find the stabilizers of a given sub space in quantum information in~\cite{houshmand2015ga}.

Realizing a quantum computer has many obstacles. There are technological limitations on the number of qubits that can be used for building a monolithic quantum computing device~\cite{van2016path}. These limitations are the main causes for the emerging of distributed quantum computing.
Grover~\cite{grover1997quantum}, Cleve and Buhrman~\cite{Cleve1997} and later Cirac et.al.~\cite{Cirac1999} were the first ones to propose distributed model of quantum computing. In~\cite{grover1997quantum}, Grover presented a distributed quantum system where there are some particles at separated locations and each one performs its computation and sends the required information to a base station when necessary. Grover showed that using this distributed approach, the overall computation time is proportional to the number of such distributed particles. Beals et.al~\cite{beals2013efficient} showed that an arbitrary quantum circuit can be emulated by a distributed quantum circuit with nodes connected using a hypercube graph. Yepez~\cite{yepez2001type} presented an architecture for distributed quantum computing with two types of communication which were called Type I and Type II. A Type-I quantum computer uses quantum communication between subsystems and Type-II quantum computer exploits classical communication between the subsystems of the distributed computer. In Type-I quantum computers, each qubit can be entangled with any number of qubits. On the other hand, a Type-II quantum computer is a network of small quantum systems and communication between them is accomplished by means of classical communication channels.

Related to teleportation cost, Streltsov et.al,~\cite{Streltsov2012} posed the question of the cheapest way for distributing entanglement and provided the minimal quantum cost for sending an entangled composite state in a long distance. They showed that regarding the most general distribution protocol, the amount of entanglement sent in the total process of distributed communication cannot be larger than the total cost for sending the ancillary particle and sending back that particle. Classical communication cost is the subject of the study in~\cite{lo2000classical}. Authors then conjectured that in a two-stage teleportation, each step requires a single bit of classical communication and in general, for an arbitrary $N$-dimensional pure state, $2\,\log \,2N$ bits of classical communications is required for remote preparation which is different from the usual teleportation in which the precise state of the qubit to be prepared in the receiver is known to the sender. Ying and Feng~\cite{ying2009} provided some definitions of a distributed quantum computing system and defined an algebraic language to describe quantum circuits for distributed quantum computing.

In~\cite{meter12008}, a distributed circuit model of a monolithic quantum circuit has been presented. In that study, the functionality of a 2-qubit VBE adder is distributed over two nodes. The adder has been split into two almost equal quantum circuits and the communication between these nodes was performed using teleportation. The authors have not considered multiple issues regarding the distributed version of the monolithic circuit. The first issue is about finding the minimal number of qubits for teleporting. The presented idea in that study works for small circuits, however, for large quantum circuits, finding the qubits for teleportation that leads to the minimal quantum cost is an important challenge. Another issue in~\cite{meter12008} is about having some consecutive gates in the target node that use the teleported qubit as one of their inputs. In their sample case, the teleported qubit is used consecutively in two gates of the destination node, but this is not always the case. Sometimes the teleported qubit is required in the source node and so somewhere in the middle the qubit should be returned back by another teleportation and this will lead to larger distributed circuits and cost in the system.

Another distributed implementation of a quantum circuit can be found in~\cite{yimsiriwattana2004distributed} where the distributed quantum circuit for Shor's algorithm using non-local gates has been implemented. The additional overhead for this distributed quantum circuit has been calculated in terms of the number of teleportation circuits, but no attempt has been done to reduce the number of teleportations or to justify that the mentioned circuit is minimaml in terms of teleportation circuits.

In our previous study~\cite{zomorodi2018optimizing}, with two given quantum subsystems, an algorithm with an exponential complexity was proposed to optimize the number of qubit teleportations required for the communication between these two subsystems. Reducing the problem of quantum circuit partitioning to the graph partitioning model has been proposed in~\cite{andresautomated}. The authors have mapped the quantum circuit into a hypergraph and then they have used the state-of-the-art hypergraph partitioning algorithms for partitioning the graph into multiple quantum circuits.
The authors have considered the case where different quantum gates have a common control or common target and then for each case they have built a hyperedge connecting common qubits and non-common qubits which meet at one end. But the authors have not considered any optimization like moving gates back and forth for making them close to each other as the approach presented in~\cite{zomorodi2018optimizing}. They have not also taken into account the entire search space of different partitioning and different partition for executing global gates and hence cannot produce optimal solutions.
\section{Proposed Approach}
\label{sec:proposed}
In this section, the proposed approach for minimizing the number of teleportations in a distributed quantum circuit is presented.
\subsection{Problem Definition}
A Distributed Quantum Circuit (DQC) which is also called a quantum multicomputer~\cite{meter12008,meter2006architectural}, consists of $N$-limited capacity Quantum Circuits (QCs) or partitions which physically are located far from each other and altogether emulate the functionality of a large quantum system. Partitions of a DQC communicate by sending their qubits to each other. In each partition, qubits are numbered from top to bottom where the $i^{th}$ line represents the $i^{th}$ qubit.

We intend to start with a quantum circuit $QC$, composed of basic gate library~\cite{houshmand2014}, i.e., CNOT and single-qubit gates, already split into two partitions. It is already known~\cite{Shende06,houshmand2014,houshmand2017quantum} that arbitrary $n$-qubit quantum gates can be decomposed to the basic gate library. We define two types of CNOT gates, namely, local and global gates. A \emph{local} CNOT gate is the one whose control and target qubits belong to the same partition. A \emph{global} CNOT gate is the one whose control and target qubits belong to different partitions.

The partition to which each qubit $q$ of a global CNOT gate belongs is called the \emph{home} partition of $q$. In order to perform a global CNOT gate, one of its two qubits should be teleported from its home partition to another. This qubit is called a \emph{migrated} qubit, as long as it is not teleported back to its home partition.

It is supposed that local gates including single-qubit and local CNOT gates are performed in their local partitions. The total number of gates in a $QC$ and the number of global gates are denoted by $m_t$ and $m_g$ respectively.

Here two questions arise:
\begin {itemize}
\item
When a global CNOT gate is supposed to be executed, which qubit of that global gate, namely the control or the target qubit, should be teleported to the other system? Does the answer influence the teleportation cost?
\item
Another question is when the teleported qubit should be returned back to its home partition? Does the answer influence the teleportation cost?
\end {itemize}

The answer to the first question is related to the second one. Therefore, we first address the second question. It is clear that when a qubit is transferred into another partition, it no longer exists in its home partition, so it should be returned to its home partition for local gate executions. In this regard, the existing works such as~\cite{meter12008} assume that as soon as an operation is applied to the teleported qubit, it returns back by another teleportation.

In order to perform global CNOT gates, both their qubits should exist in the same partition. Teleportation is the natural way of transporting qubit states. In a two-partite system, with Partitions $A$ and $B$, there are two cases for executing each gate. One is the teleportation of the qubit in Partition $A$ to Partition $B$ and then executing the gate in Partition $B$ and vice versa.

In~\cite{zomorodi2018optimizing}, an exact solution has been given and an algorithm has been proposed to minimize the number of teleportations in a two-partite distributed quantum circuit. The approach calculates the minimal number of teleportations for each configuration of global gates, where each configuration has a unique placement for global gates being in either Partition $A$ or $B$. Finally, the minimal number of teleportations for all of the configurations is returned. For a two-partite system with $m_g$ global gates, there are $2^{m_g}$ different configurations of executing global gates whose consideration takes an exponential complexity of $O(2^{m_g})$.

In order to overcome this problem, in this paper, a heuristic approach based on genetic algorithms is proposed which attempts to find a configuration of global gates with an optimized teleportation cost. Each chromosome, i.e., the individuals in the GA population of potential solutions, shows a configuration whose fitness is computed by \emph{MIN-TELEPORTATION} function in~\cite{zomorodi2018optimizing} and then by the evolution in the generations, the configuration with the optimized cost is found. The details of the proposed approach are given in the following sections.
%%In the \emph{MIN-TELEPORTATION} function, when a qubit of a global gate is teleported to the other partition,
%the whole circuit is tracked and as much as possible number of gates that can be executed without the need of teleporting this qubit back are executed. This implies that the migrated qubit has been used optimally by other gates before it is teleported back to its own partition.

\subsection{The outline of the algorithm}
In the following, the details of the proposed genetic algorithm are presented.

\subsubsection{Encoding of chromosomes and initial population}
\begin{figure}[!htb]
\begin{center}
\includegraphics[width=0.4\textwidth]{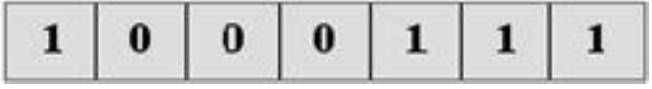}
\end{center}
\caption{The representation of chromosomes}
\label{fig1}
\end{figure}

One of the most important parts in GA is the encoding mechanism for representing chromosomes. In this paper, the algorithm uses a chromosome structure with $m_g$ genes which is basically a string of $m_g$ bits where $m_g$ is the number of global gates. The $i^\text{{th}}$ gene corresponds to the $i^\text{{th}}$ global gate, where the gates are labeled from left to right. If a global gate is considered to be executed in Partition $A$ ($B$), the corresponding gene takes the value of 0 (1) respectively.
For example, for Figure~\ref{fig1}, $m_g$ is equal to 7. It is inferred that the first global gate is executed in Partition $B$, the second global gate is intended for execution in Partition $A$, and so on. The chromosome illustrates the global gate configuration in the distributed circuit. Each configuration can be a potential solution to the problem. The algorithm starts with a random initial population of chromosomes.

\subsubsection{Fitness function and selection strategy}
The fitness function for this genetic algorithm is computed as the minimal number of required teleportations for a given configuration of global gates, represented by chromosomes, of a distributed quantum system. This value is computed by the \emph{MIN-TELEPORTATION} function in~\cite{zomorodi2018optimizing}. In the \emph{MIN-TELEPORTATION} function, when a qubit of a global gate is teleported to the other partition,
the whole circuit is tracked and as much as possible number of gates that can be executed without the need of teleporting this qubit back are executed. This
implies that the migrated qubit has been used optimally by other gates before it is teleported back to its own partition.
The \emph{MIN-TELEPORTATION} function is a recursive function, which basically includes two other functions, called, \emph{NON-EXECUTE}, and \emph{NON-COMMUTE}. The \emph{NON-EXECUTE} function takes two gates, $g$ and $g'$ as inputs. It returns TRUE, if the
migrated qubit of $g$ should be returned to its home partition and then $g'$ can be performed. It returns FALSE otherwise. $NON$-$COMMUTE(g, g')$ takes two gates, $g$ and $g'$ as inputs. It returns TRUE,
if the gates $g$ and $g'$ do not commute and returns FALSE otherwise. For more details about these functions, the readers are referred to~\cite{zomorodi2018optimizing}. The roulette wheel strategy is used for the selection.
%without the need of teleporting back the qubit are executed. This means that the migrated

\subsubsection{Crossover, mutation, and replacement}
Crossover recombines two randomly selected ancestors into two fresh off-spring. We used two-point crossover where two crossover points on both parents' chromosomes are randomly selected. Then, the data between those points in either chromosome are swapped between the two parents'
chromosomes. Mutation is implemented by randomly inverting a randomly chosen gene of a selected individual. The random numbers generated for choosing ancestors, genes and changing the genes use a uniform distribution.
Replacement forms the next generation of individuals by replacing some offspring using a specific replacement strategy. In this study, we replaced a number of the worst chromosomes with the best ones which are carried over to the next generation unaltered. This strategy is known as elitism which guarantees that the solution quality obtained by the GA will not decrease from one generation to the next. Mutation, crossover, and replacement take place with their corresponding probabilities, i.e., $P_m$, $P_c$, and $P_r$, respectively.

\section{Results}
\label{sec:results}
We implemented our algorithm in MATLAB on a workstation with 4 GB RAM and 2.0 GHz CPU to find the configuration with an optimized number of teleportations.
According to the complexity of the problem in this study, the wrong choice of GA parameters might greatly increase the convergence time of GA. It is explained in the following how the GA parameters are set.

The number of initial population is one of the important parameters in evolutionary algorithms like GA. There
have been many studies~\cite{rylander2002optimal,he2002individual,oliveto2009analysis} analyzing the impact of different population sizes on the performance of GA. The basic idea is that a larger population size may increase the population diversity and consequently help GA, however, it is shown~\cite{chen2012large} that there are conditions where larger populations might be harmful.

On the other hand, it is already known that in any search algorithm, including GA, we seek a proper balance between exploration and exploitation~\cite{vcrepinvsek2013exploration}. Exploration is the process of visiting entirely new zones of a search space,
while exploitation is the process of visiting those zones of a search space within the neighborhood of previously visited points. In GA, the mutation operator is mostly used to provide exploration so as to increase the probability of finding the optimal solution while the crossover operator is widely used to lead population to converge to the optimal solution (exploitation). This balance is determined by the mutation and the crossover rate. Moreover, the replacement strategy as used in this study (elitism) can help to keep best chromosomes in all generations.

As the parameters of GA depend on the problem, we tuned these parameters by performing different experiments and have set parameters such that they produce a better solution within a certain amount of time. The values of the GA parameters are given in Table~\ref{Table1}.
\begin{table}
\centering
  \caption{Parameters of GA. $P_m$, $P_c$, and $P_r$ denote probabilities of mutation, crossover and replacement, respectively. $m_g$ denotes the number of global gates in the given quantum circuit.}
\label{Table1}

\begin{tabular}{|l|l|l|l|l|}

\hline
\# of initial population  & $P_m$ & $P_c$ &  $P_r$\\ \hline
$\left\lceil {\frac{{m_g }}{2}} \right\rceil$ & 0.1  & 0.9 & 0.4 \\ \hline
\end{tabular}
\end{table}
We stopped GA if $a)$ the maximum number of 1000 generations is reached or $b)$ the improvement in the fitness value of the best individual in the population over 10 generations is less than 0.001.

For comparing the performance of our algorithm with the exact solution of~\cite{zomorodi2018optimizing}, we used some circuits from Revlib~\cite{wille2008revlib} library which is an online resource for benchmarks within the domain of reversible and quantum circuits, and the other set is QFT($n$) where $n \in \{4,8,16,32,64\}$. %We tested QFT for five different $n$s.

The benchmark circuits were first decomposed into the basic gate library by applying the synthesis approach in~\cite{Barenco95,Shende06}.
The proposed approach should first assign qubits to two partitions. To this end, the initial partitioning of the circuits into two partitions is performed by the Kernighan-Lin (K-L) algorithm as used in the VLSI design algorithms~\cite{kernighan1970efficient}. The K-L algorithm
is a heuristic algorithm for solving the graph partitioning problem. We use it to partition a graph into two partitions in such a way the interconnection cost
between different partitions is minimized. The complexity of the K-L algorithm is $O(n^2log(n))$ where $n$ is the number of nodes in the graph which is equal to the number of qubits in our problem.

We compared the results of the proposed approach with~\cite{zomorodi2018optimizing}. Moreover, in order to show the effectiveness of GA, we performed a random search over different configurations and found the minimal number of required teleportations in all of the iterations. The number of iterations of the random search was equal to the size of initial population in GA multiplied by the number of generations of our GA when it was terminated.

\begin{figure}[!ht]
  \centering
    \includegraphics[width=1\textwidth]{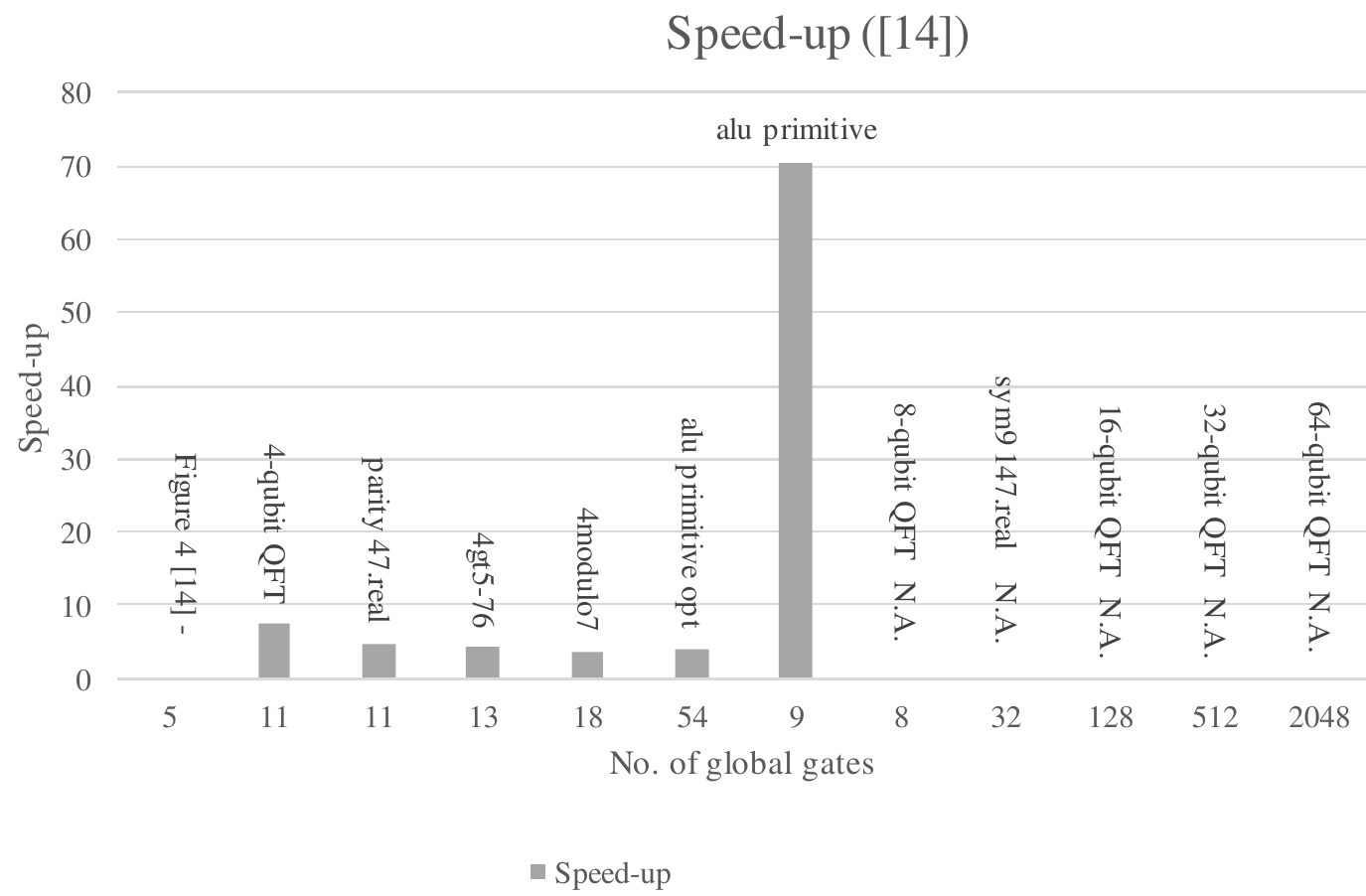}
  \caption{Speed-up of the proposed approach as compared to~\cite{zomorodi2018optimizing}}
  \label{fig:speedup}
\end{figure}

\begin{figure}[!ht]
  \centering
    \includegraphics[width=1\textwidth]{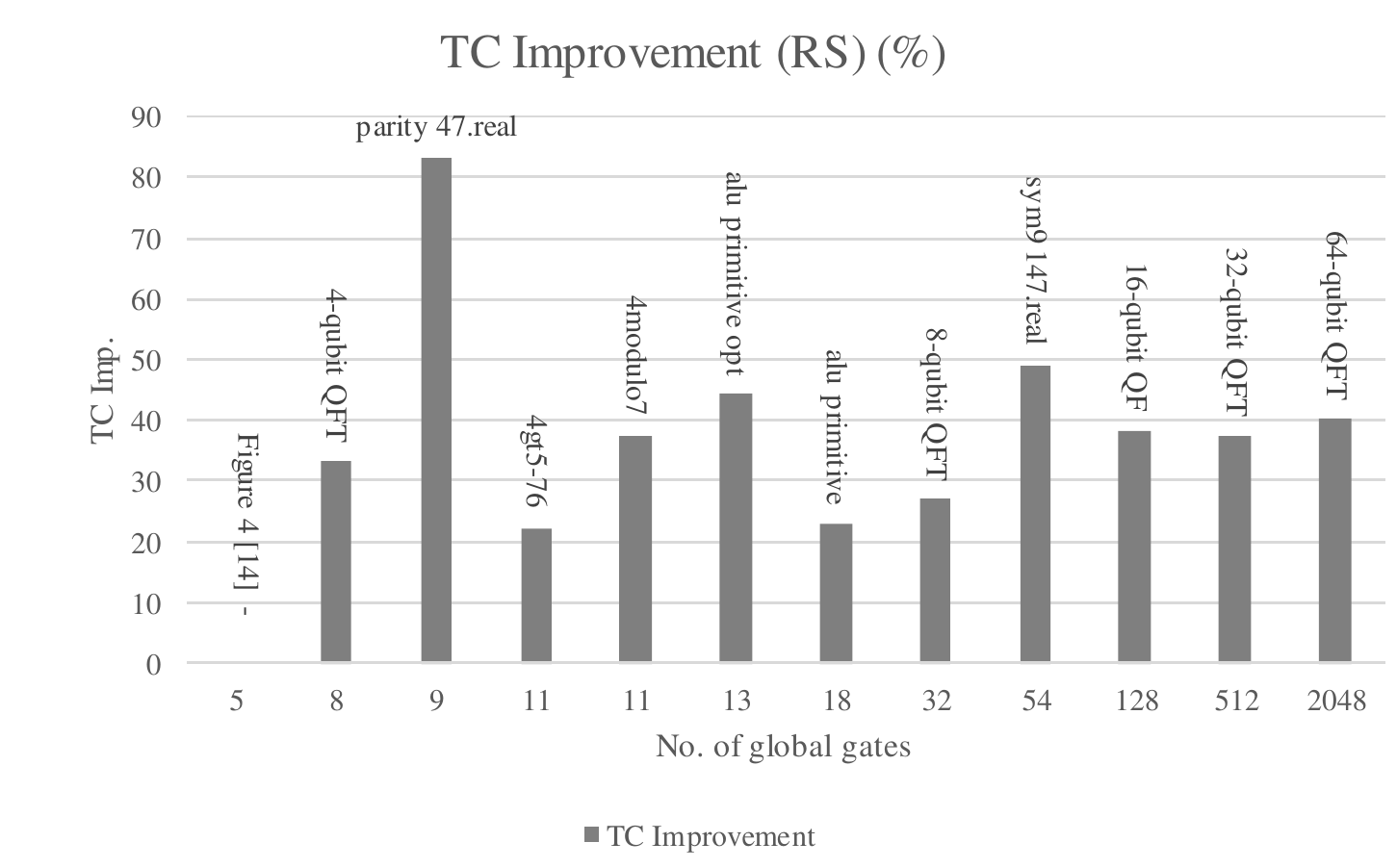}
  \caption{Percentage of the teleportation cost (TC) improvement of the proposed approach as compared to the random search (RS) }
  \label{fig:tcimprovement}
\end{figure}

Table~\ref{table:table1} compares the results of our algorithm with~\cite{zomorodi2018optimizing} and the random search in terms of the execution time and the teleportation cost (TC), respectively, for 12 different benchmark circuits. In Figures~\ref{fig:speedup} and~\ref{fig:tcimprovement} the speed-up of the proposed approach as compared to~\cite{zomorodi2018optimizing} and the percentage of the TC improvement of the proposed approach as compared to the random search for different benchmarks of Table~\ref{table:table1} are shown respectively. The horizontal axis in these figures relates to the number of global gates as the search space size of~\cite{zomorodi2018optimizing} is exponential in terms of the number of global gates of the circuits.

It should be noted that for the small circuits, e.g., Figure 4 of~\cite{zomorodi2018optimizing}, where there are a few global gates, GA takes more time due to its overhead
and there is no gain in applying GA. However, when the number of global gates
increases, GA shows more advantage by remarkable speed-ups.  For the cases in this table where the number of global gates is equal to or greater than 32, the approach of \cite{zomorodi2018optimizing} cannot produce the answer even after ten days of continuous running, which are reported as N.A. The resultant
TCs for the proposed approach and the approach of~\cite{zomorodi2018optimizing} for the benchmark
circuits (excluding the N.A. rows) are the same except for alu-primitive (with 18 global gates), even though GA has a considerably
lower execution time. The approach of~\cite{zomorodi2018optimizing} computes the teleportation cost of 18 for alu-primitive while the proposed reaches to the teleportation cost of 20 for this benchmark which are only slightly different. A speed-up of 70.54 is obtained for this benchmark circuit. It is concluded that the genetic algorithm is capable of producing almost the same results with the optimal solution of \cite{zomorodi2018optimizing} in much less time.

The proposed approach also improves the teleportation cost of the purely random search by on average $36.33\%$ for the benchmark circuits which verifies that by taking inspirations from natural evolution concepts, GA can converge to near-optimal results. For the parity 47.real circuit which has a regular structure, the proposed approach can effectively produce a TC of 2, while the random search cannot find the optimal answer and there a is $83.33\%$ improvement in the TC for this circuit.

\begin{table}[!ht]
\caption{Comparison of the proposed approach (P) with the random search (RS) and~\cite{zomorodi2018optimizing}. The last two columns indicate the percentage of teleportation cost improvement (TC. imp) and the speed-up of the proposed approach as compared to RS and~\cite{zomorodi2018optimizing}, respectively.}
\label{Table2}
\scalebox{0.9}{
\begin{tabular}{llllllllll}
Circuit             & \begin{tabular}[c]{@{}l@{}}\# of\\  qubits\end{tabular} & \begin{tabular}[c]{@{}l@{}}\# of \\ global\\  gates\end{tabular} & \begin{tabular}[c]{@{}l@{}}TC\\ (\cite{zomorodi2018optimizing})\end{tabular} & \begin{tabular}[c]{@{}l@{}}TC \\ (RS)\end{tabular} & \begin{tabular}[c]{@{}l@{}}TC \\ (P)\end{tabular} & \begin{tabular}[c]{@{}l@{}}Time (S)\\  (\cite{zomorodi2018optimizing})\end{tabular} & \begin{tabular}[c]{@{}l@{}}Time \\ (S) (P)\end{tabular} & \begin{tabular}[c]{@{}l@{}}TC. \\ imp \\ (RS)\\ (\%)\end{tabular} & \begin{tabular}[c]{@{}l@{}}Speed-up \\ (\cite{zomorodi2018optimizing})\end{tabular} \\
Figure 4~\cite{zomorodi2018optimizing}            & 4                                                       & 5                                                                & 4                                                     & 4                                                  & 4                                                 & 0.65                                                         & 3.15                                                    & -                                                                & -                                                            \\
4gt5-76             & 5                                                       & 11                                                               & 14                                                    & 18                                                 & 14                                                & 1194.92                                                      & 275.88                                                  & 22.22                                                            & 4.17                                                         \\
4modulo7            & 5                                                       & 11                                                               & 10                                                    & 16                                                 & 10                                                & 1013.36                                                       & 292.26                                                  & 37.5                                                             & 3.46                                                         \\
alu\_primitive\_opt & 6                                                       & 13                                                               & 10                                                    & 18                                                 & 10                                                & 4874.24                                                      & 1182.51                                                 & 44.44                                                            & 4.12                                                         \\
alu\_primitive      & 6                                                       & 18                                                               & 18                                                    & 26                                                 & 20                                                & 154020.53                                                    & 2183.31                                                 & 23.06                                                            & 70.54                                                        \\
sym9\_147.real      & 12                                                      & 54                                                               & N.A.                                                  & 94                                                 & 48                                                & N.A.                                                         & 10930.57                                                & 48.93                                                            & N.A.                                                         \\
parity\_47.real     & 17                                                      & 9                                                                & 2                                                     & 12                                                 & 2                                                 & 1028.63                                                      & 212.78                                                  & 83.33                                                            & 4.83                                                         \\
4-qubit QFT         & 4                                                       & 8                                                                & 8                                                     & 12                                                 & 8                                                 & 875.62                                                       & 170.13                                                  & 33.33                                                            & 7.49                                                         \\
8-qubit QFT         & 8                                                       & 32                                                               & N.A.                                                  & 52                                                 & 38                                                & N.A.                                                         & 8059.40                                                & 26.92                                                            & N.A.                                                         \\
16-qubit QFT        & 16                                                      & 128                                                              & N.A.                                                  & 216                                                & 133                                               & N.A.                                                         & 19788.34                                                & 38.42                                                            & N.A.                                                         \\
32-qubit QFT        & 32                                                      & 512                                                              & N.A.                                                  & 852                                                & 532                                               & N.A.                                                         & 78734.89                                               & 37.55                                                            & N.A.                                                         \\
64-qubit QFT        & 64                                                      & 2048                                                             & N.A.                                                  & 3752                                               & 2250                                              & N.A.                                                         & 361645.67                                               & 40.31                                                            & N.A.
\end{tabular}}
\label{table:table1}
\end{table}

\section{Conclusion}
\label{conclusion}
This paper proposed an approach based on the genetic algorithm to obtain a configuration of global gates in a distributed quantum circuit which leads to minimal number of teleportation. Compared with previous work in~\cite{zomorodi2018optimizing}, with an exponential complexity, it was shown that the present proposed work yielded almost the same results in much less time. Moreover, the results demonstrate that GA decreases teleportation cost by on average $36.33\%$ as compared with a random search over configurations and verified the effectiveness of GA as a guided random search.
%We conclude that by using this algorithm the best configuration of global gates can be achieved.
As future works, we are going to consider the case where the number of partitions is more than two. There are some challenges for having more than two partitions including the representation of global gates and finding the best configuration.
Also, interwinding the initial partitioning scheme of qubits into subcircuits with the phase of obtaining an optimized configuration of global gates can be considered as a future work.

\bibliographystyle{unsrt}
\bibliography{ref}
\end{document}